\documentclass[aps,prb,superscriptaddress,twocolumn,floatfix]{revtex4-1}
\usepackage{graphicx}
\usepackage{amssymb}
\usepackage{amsfonts}
\usepackage{amsmath}

\begin{document}

\title{Muon spin rotation study of the topological superconductor Sr$_x$Bi$_2$Se$_3$}

\author{H. Leng} \email{h.leng@uva.nl}\affiliation{Van der Waals - Zeeman Institute, University of Amsterdam, Science Park 904, 1098 XH Amsterdam, The Netherlands}
\author{D. Cherian} \affiliation{Van der Waals - Zeeman Institute, University of Amsterdam, Science Park 904, 1098 XH Amsterdam, The Netherlands}
\author{Y. K. Huang} \affiliation{Van der Waals - Zeeman Institute, University of Amsterdam, Science Park 904, 1098 XH Amsterdam, The Netherlands}
\author{J.-C. Orain} \affiliation{Laboratory for Muon-Spin Spectroscopy, Paul Scherrer Institute, 5232 Villigen PSI,  Switzerland}
\author{A. Amato} \affiliation{Laboratory for Muon-Spin Spectroscopy, Paul Scherrer Institute, 5232 Villigen PSI,  Switzerland}
\author{A. de Visser} \email{a.devisser@uva.nl} \affiliation{Van der Waals - Zeeman Institute, University of Amsterdam, Science Park 904, 1098 XH Amsterdam, The Netherlands}

\date{\today}

\begin{abstract}
We report transverse-field (TF) muon spin rotation experiments on single crystals of the topological superconductor Sr$_x$Bi$_2$Se$_3$ with nominal concentrations $x=0.15$ and $0.18$ ($T_c \sim 3$~K). The TF spectra ($B= 10$~mT), measured after cooling to below $T_c$ in field, did not show any additional damping of the muon precession signal due to the flux line lattice within the experimental uncertainty. This puts a lower bound on the magnetic penetration depth $\lambda \geq 2.3 ~\mu$m. However, when we induce disorder in the vortex lattice by changing the magnetic field below $T_c$ a sizeable damping rate is obtained for $T \rightarrow 0$. The data provide microscopic evidence for a superconducting volume fraction of $\sim 70~\%$ in the $x=0.18$ crystal and thus bulk superconductivity.

\end{abstract}


\maketitle

\section{INTRODUCTION}

Sr$_x$Bi$_2$Se$_3$ belongs to the new family of Bi$_2$Se$_3$-based superconductors, which is reported to exhibit unconventional superconducting properties. The parent compound Bi$_2$Se$_3$ is a well documented, archetypal topological insulator~\cite{Zhang2009,Xia2009,Hsieh2009}. Recently, it was demonstrated that by doping Cu~\cite{Hor2010}, Sr~\cite{Liu2015}, Nb~\cite{Qiu2015} or Tl~\cite{Wang2016} atoms Bi$_2$Se$_3$ can be transformed into a superconductor with $T_c \sim 3$~K. Theory predicts the superconducting state to have a topological character, which is based on the close correspondence of the Bogoliubov - de Gennes Hamiltonian for the quasiparticles of the superconductor and the Bloch Hamiltonian for the insulator (for recent reviews on topological superconductivity see Refs~\onlinecite{Mizushima2016,Sato&Ando2017}). In a topological superconductor the condensate is expected to consist of Cooper pairs with odd parity symmetry, while at the surface of the material gapless Andreev bound states form that host Majorana zero modes. This provides an excellent motivation to thoroughly examine the family of Bi$_2$Se$_3$-based superconductors. These centrosymmetric compounds ($D_{3d}$ point group, $R \overline{3}m$ space group) belong to the symmetry class DIII~\cite{Schnyder2008}. Calculations within a two-orbital model show that odd-parity pairing, favoured by strong spin-orbit coupling, can be realized~\cite{Fu&Berg2010}. In the case of Cu$_x$Bi$_2$Se$_3$ specific heat~\cite{Kriener2011a}, upper critical field~\cite{Bay2012b} and soft-point contact experiments~\cite{Sasaki2011} lend support to an odd parity superconducting state. However, scanning tunneling microscopy (STM) measurements were interpreted to be consistent with $s$-wave pairing symmetry~\cite{Levy2013}. Clearly, further studies are required to solve this issue.

Superconductivity in Sr$_x$Bi$_2$Se$_3$ was discovered by Liu \textit{et al.}~\cite{Liu2015}. Transport and magnetic measurements on Sr$_x$Bi$_2$Se$_3$ single crystals with $x=0.06$ show $T_c = 2.5$~K. The resistivity is metallic with a low carrier concentration $n \approx 2 \times 10^{25}$~m$^{-3}$. Evidence for topological surface states was extracted from Shubnikov - de Haas oscillations observed in large magnetic fields~\cite{Liu2015}. The persistence of topological surface states upon Sr doping was confirmed by angle resolved photoemisison experiments (ARPES) measurements, that showed a topological surface state well separated from the bulk conduction band~\cite{Han2015,Neupane2015}. The superconducting state was further characterized by Shruti \textit{et al.}~\cite{Shruti2015} who reported $T_c = 2.9$~K for $x=0.10$ and a large Ginzburg-Landau parameter, $\kappa \approx 120$, pointing to extreme type II superconducting behavior. A surprising discovery was made by Pan \textit{et al.}~\cite{Pan2016} by performing magnetotransport measurements on crystals with nominal concentrations $x=0.10$ and 0.15: the angular variation of the upper critical field, $B_{c2} (\theta)$, shows a pronounced \textit{two-fold anisotropy for field directions in the basal plane}, i.e. the rotational symmetry is broken. Magnetotransport measurements under high pressures show the two-fold anisotropy is robust up to  at least $p=2.2$~GPa\cite{Nikitin2016}.

Most interestingly, rotational symmetry breaking appears to be a common feature of the Bi$_2$Se$_3$-based superconductors when the dopant atoms are intercalated. In Cu$_x$Bi$_2$Se$_3$ it appears in the spin-system below $T_{c}$ as was established by the angular variation of the Knight shift measured by nuclear magnetic resonance (NMR)~\cite{Matano2016}. Moreover, specific heat measurements show the basal-plane anisotropy in $B_{c2}$ is a thermodynamic bulk feature~\cite{Yonezawa2017}. In Nb$_x$Bi$_2$Se$_3$ rotational symmetry breaking was demonstrated by torque magnetometry that probes the magnetization of the vortex lattice~\cite{Asaba2017}. These recent experiments put important constraints on the superconducting order parameter. Notably, it restricts the order parameter to an odd-parity two-dimensional representation, $E_u$, with $\Delta _4$-pairing~\cite{Nagai2012,Fu2014,Venderbos2015}. Moreover, the superconducting state involves a nematic director that breaks the rotational symmetry when pinned to the crystal lattice, hence the label nematic superconductivity. The odd-parity Cooper pair state implies these Bi$_2$Se$_3$-derived superconductors are topological superconductors.

Here we report a muon spin rotation study on Sr$_x$Bi$_2$Se$_3$. The aim of our experiments was to investigate whether rotational symmetry breaking extends to deep in the superconducting phase through  measurements of the basal-plane anisotropy of the London penetration depth, $\lambda$. Muon spin rotation is an outstanding technique to determine the temperature variation of $\lambda$, as well as its absolute value, via the Gaussian damping rate, $\sigma_{TF}$, of the $\mu ^+$ precession signal in a transverse field experiment. Below $T_c$, an increase of $\sigma_{TF}$ is expected because the muon senses the additional broadening of the field distribution due to the flux line lattice~\cite{Amato1997,Blundell1999}. The measurements show, however, that the increase of $\sigma_{TF}$ is smaller than the experimental uncertainty in field-cooling experiments, which tell us $\lambda$ is very large ($\geq 2.3$~$\mu$m for $T \rightarrow 0$). On the other hand, when we induce disorder in the vortex lattice by changing the magnetic field below $T_c$ a sizeable damping rate $\sigma_{SC} \approx 0.36$~$\mu$s$^{-1}$ ($T \rightarrow 0$) is obtained. These results provide microscopic evidence for a superconducting volume fraction of $\sim 70~\%$ in the crystal with nominal Sr content $x=0.18$ and thus bulk superconductivity.

\section{EXPERIMENTAL}

Single crystalline samples Sr$_{x}$Bi$_{2}$Se$_{3}$ with nominal values $x=0.15$ and $x=0.18$, were synthesized by melting high-purity elements at 850~$^{\circ}$C in sealed evacuated quartz tubes. The crystals were formed by slowly cooling to 650~$^{\circ}$C at a rate of 3~$^{\circ}$C/hour. Powder X-ray diffraction confirmed the R$\bar{3}$m space group. The single-crystalline nature of the crystals was checked by Laue back-reflection. Thin (thickness 0.4~mm) flat rectangular crystals were cut from the single-crystalline batch by a scalpel and/or spark erosion. The sample plane contains the trigonal basal plane with the $a$ and $a^*$ axes. The sample area for the incident muon beam is $8 \times 12$~mm$^{2}$ and $3 \times 10$~mm$^{2}$ for $x=0.15$ and $x=0.18$, respectively. The characterization of the single-crystalline batch with $x=0.15$ is presented in Ref.~\onlinecite{Pan2016}. Ac-susceptibility measurements  show a superconducting shielding fraction of 80~\%. For the $x=0.18$ batch we obtain a slightly lower screening fraction, 70~\%.

Muon spin rotation ($\mu$SR) experiments were carried out with the Multi Purpose Surface Muon Instrument DOLLY installed at the $\pi $E1 beamline at the S$\mu$S facility of the Paul Scherrer Institute. The technique uses spin-polarized muons that are implanted in a sample. If there is a local or applied field at the sample position the muon spin will precess around the field direction. The subsequent asymmetric decay process is monitored by counting the emitted positrons by scintillation detectors that are placed at opposite directions in the muon-spin precession plane~\cite{Amato1997,Blundell1999}. The parameter of interest is the muon spin asymmetry function, $A(t)$, which is determined by calculating $A(t) =(N_1 (t) - \alpha N_2 (t))/(N_1 (t) + \alpha N_2 (t) )$, where $N_1 (t)$ and $N_2 (t)$ are the positron counts of the two opposite detectors, and $\alpha$ is a calibration constant. In our case $\alpha$ is close to 1. In the transverse field (TF) configuration the damping of the muon spin precession signal is a measure for the field distribution sensed by the muon at its localization site. For a superconductor below $T_c$, in a small TF of typically 10~mT, the vortex lattice is expected to produce a Gaussian damping, $\sigma_{SC} = \gamma_{\mu} \sqrt{\langle
(\Delta B)^2 \rangle}$, with $\gamma_{\mu} = 2 \pi \times 135.5~$MHz/T the muon gyromagnetic ratio and $\langle (\Delta B)^2 \rangle$ the second moment of the field distribution. TF experiments were performed for a field along the $a$-axis and the $c$-axis. In the first case the muon spin is horizontal, i.e. along the beam direction, and the positrons are collected in the forward and backward detectors. In the second case the muon spin is vertical (spin rotated mode), the field is applied along the beam, and the positrons are collected in the left and right detectors. The crystals were glued with General Electric (GE) varnish to a thin copper foil, that was attached to the cold finger of a helium-3 refrigerator (HELIOX, Oxford Instruments). $\mu$SR spectra were taken in the temperature interval $T=0.25 - 10$~K. The $\mu$SR time spectra were analysed with the software packages WIMDA~\cite{Pratt2000} and Musrfit~\cite{Suter&Wojek2012}.

\section{RESULTS AND ANALYSIS}

\subsection{Field-cooled spectra}

\begin{figure}[ht]
\centering
\includegraphics[width=7cm]{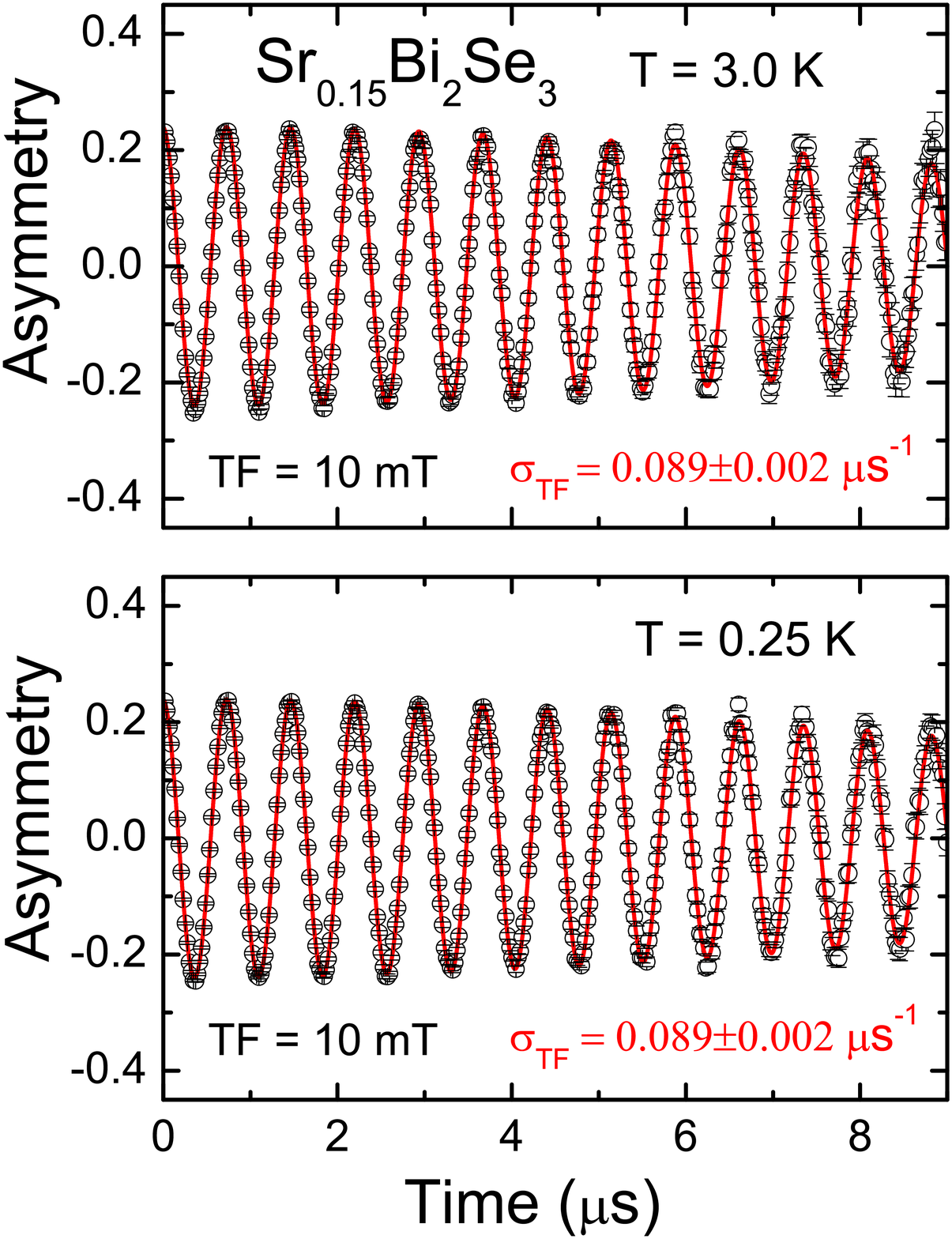}
\caption{$\mu$SR spectra for Sr$_{0.15}$Bi$_{2}$Se$_{3}$ measured in a transverse field of 10~mT ($B \parallel a)$ at $T=3.0$~K (upper panel) and $T=0.25$~K (lower panel). The red lines are fits using the muon depolarization function Eq.~1. The spectra are taken after field cooling in 10~mT.}
\label{figure:TFspectra}
\end{figure}

A first set of experiments was carried out for $x=0.15$. The crystal with $T_c = 2.8$~K was slowly cooled in a TF field of 10~mT ($B \parallel a$) to $T=0.25$~K, after which $\mu$SR spectra were recorded at fixed temperatures, during step-wise increasing the temperature up to 3.0~K. The measured spectra at 0.25~K and 3.0 K are shown Fig.~\ref{figure:TFspectra}, where we have plotted the decay asymmetry as a function of time. The initial asymmetry $A(0) = 0.24$ is the full experimental asymmetry ($A_{tot}$). As can be noticed, the spectra at 0.25~K and 3.0~K are very similar. We have fitted the spectra with the muon depolarization function
\begin{equation}
  A (t)= A_{tot} \exp (-\frac{1}{2} \sigma_{TF}^2 t^2) \cos (2\pi\nu t+\phi) .
\end{equation}

\begin{figure}[ht]
\centering
\includegraphics[width=7cm]{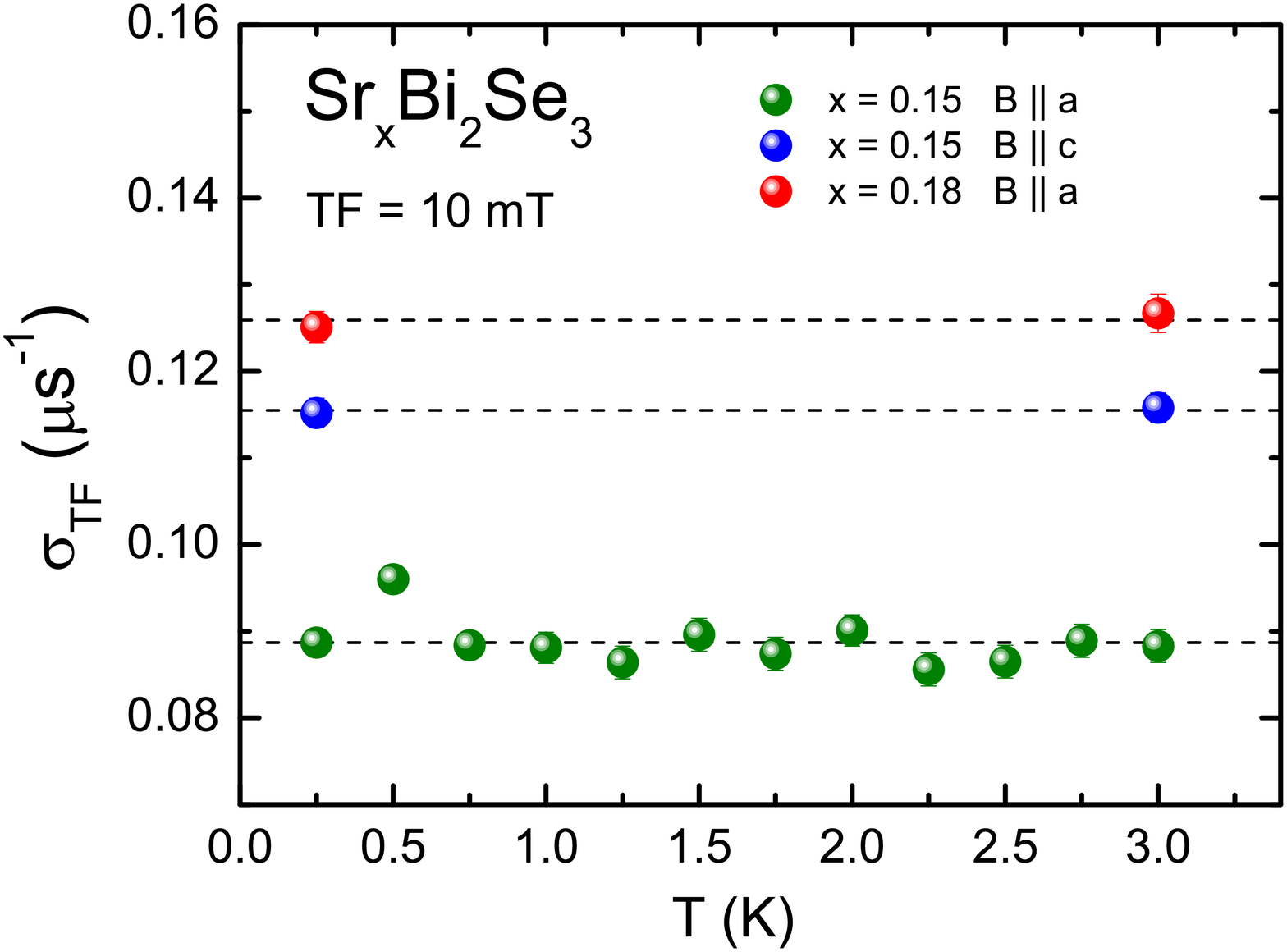}
\caption{Temperature variation of the Gaussian damping rate $\sigma _{TF}$ extracted with help of Eq.~1 from TF $\mu$SR spectra taken after field cooling in 10~mT. Green symbols: for $x=0.15$ and $B \parallel a$. Blue symbols: for $x=0.15$ and $B \parallel c$. Red symbols:  for $x=0.18$ and $B \parallel a$. The dashed horizontal lines show $\sigma_{TF}$ is temperature independent.}
\label{figure:sigmaTF}
\end{figure}

Here $\sigma _{TF}$ is the Gaussian damping rate, $\nu = \gamma_{\mu} B_{\mu} /2 \pi$ is the muon precession frequency, $B_{\mu}$ is the average field sensed by the muon ensemble and $\phi$ is a phase factor. The resulting temperature variation $\sigma_{TF} (T)$ is shown in Fig.~\ref{figure:sigmaTF}. In the normal phase $\sigma _{TF}=0.089 \pm 0.002 ~ \mu$s$^{-1}$, which we attribute to the field distribution due to nuclear moments considered static within the $\mu$SR time window. No additional damping is observed below $T_c$ within the experimental resolution and we conclude $\sigma _{SC}$ is very small. An upper bound for $\sigma _{SC}$ can be derived with help of the equation~\cite{Luke1993}
\begin{equation}
  \sigma_{SC}=(\sigma_{TF, T<T_c}^2 - \sigma_{TF, T>T_c}^2 )^{1/2}   .
\end{equation}
With the experimental uncertainty in $\sigma_{FL}$ of $\pm 0.002 ~\mu$s$^{-1}$ we obtain $\sigma_{SC} \leq 0.02 ~ \mu$s$^{-1}$. This allows us to determine a lower bound for the London penetration depth. In the vortex state of an extreme type II superconductor with a trigonal flux line lattice  $\lambda$ can be estimated from the second moment of the field distribution for $B > B_{c1}$ via the relation ${\langle
(\Delta B)^2 \rangle}=0.003706 \times \Phi_0 ^2 / \lambda^4$, where $\Phi_0$ is the flux quantum~\cite{Brandt1988a}, or
\begin{equation}
\lambda = (0.0609 \gamma_u \Phi_0 / \sigma_{SC})^{1/2}   .
\end{equation}
With $\sigma_{SC}=0.02 ~\mu$s$^{-1}$ we calculate $\lambda \geq 2.3 ~\mu$m for $T \rightarrow 0$.

In the experimental configuration used to measure the data in Fig.~\ref{figure:sigmaTF} ($B \parallel a$) we probe the penetration depths orthogonal to the field direction, or rather the product $\lambda _c \lambda _{a^*}$. We have also carried out field-cooled (10~mT) measurements for $B \parallel c$ (muon spin rotated mode; here $A_{tot}=0.19$) to probe the product $\lambda _a \lambda _{a^*}$. The extracted $\sigma_{TF}$-values at 0.25~K and 3.0~K are slightly larger than for $B \parallel a$, but are equal within the experimental resolution, as shown in Fig.~\ref{figure:sigmaTF}. Finally, we have measured field-cooled $\mu$SR spectra on the $x=0.18$ crystal for $B \parallel a$ at 0.25~K and 3.0~K. The analysis shows $\sigma_{TF} = 0.126 \pm 0.002~\mu$s$^{-1}$ and, again, no significant temperature variation in $\sigma_{TF}$ is observed as shown in Fig.~\ref{figure:sigmaTF}. We conclude, for both crystals the London penetration is very large and a conservative lower bound is $\lambda = 2.3~\mu$m.

\subsection{Vortex lattice with disorder}

The standard procedure, used above, to extract $\lambda$ from the $\mu$SR spectra for a type II superconductor relies on cooling the crystal in a small magnetic field $> B_{c1}$, which tends to produce a well ordered flux line lattice. Eq.~3 can then be used to calculate $\lambda$ once $\sigma_{SC}$ is determined~\cite{Brandt1988a}. It is well known that inducing disorder in the vortex lattice increases the distribution of the internal magnetic fields, and hence $\sigma_{SC}$~\cite{Brandt1988b,Belousov1990,Aegerter&Lee1997}. In this case, $\lambda$ can no longer be calculated with help of Eq.~3, because the calculation of $\lambda$ from the field distribution has become an intricate problem~\cite{Brandt1988b,Belousov1990,Aegerter&Lee1997}. Inducing disorder in the vortex lattice provides however an appealing route to probe the superconducting volume fraction of our crystals.

\begin{figure}[ht]
\centering
\includegraphics[width=7cm]{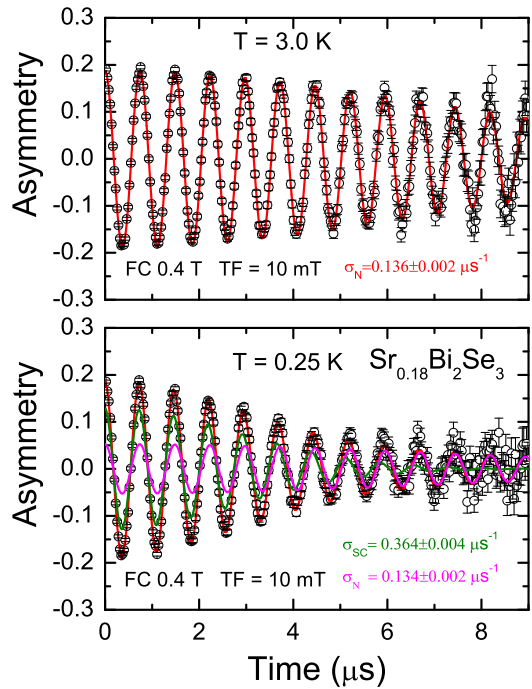}
\caption{$\mu$SR spectra for Sr$_{0.18}$Bi$_{2}$Se$_{3}$ measured in a transverse field of 10~mT at $T=3.0$~K (upper panel) and $T=0.25$~K (lower panel). The crystal was cooled in a strong field of 0.4~T ($B \parallel c$), after which the field was reduced to 10~mT. The red lines are fits to the muon depolarization function Eq.~4. The green and magenta lines in the lower panel represent the contributions to the $\mu$SR signal from the superconducting and normal state volume fractions, respectively.}
\label{figure:TFspectra_disorder}
\end{figure}

A standard procedure to induce disorder in the flux line lattice is to cool the sample to below $T_c$ in zero field and then sweep the field to the desired TF value (ZFC mode). Examples in the literature that show a pronounced increase of $\sigma_{SC}$ due to disorder can be found in Refs.~\onlinecite{Luke1993,Wu1993}. Here we followed a different procedure and cooled the $x=0.18$ crystal in a strong magnetic field ($B \parallel c$) of 0.4~T to $T=0.25$~K, after which the field was reduced to 10~mT. Decreasing the applied field causes the flux lines to move. Pinning of flux lines at crystalline defects and inhomogeneities generates magnetic disorder. We remark that for an applied field of 0.4~T the lattice parameter of the trigonal vortex lattice is $a_{\triangle}=(4/3)^{1/4}(\Phi_0 /B)^{1/2} = 0.08~\mu$m. After decreasing the field to 10~mT $a_{\triangle} = 0.49~\mu$m. Next, TF=10~mT $\mu$SR spectra were taken in the temperature range 0.25~-~5~K by step-wise increasing the temperature. In Fig.~\ref{figure:TFspectra_disorder} we show the data taken at 0.25~K and 3.0~K. As expected, a pronounced damping now appears in the superconducting state. We first fitted the spectrum at 0.25~K to Eq.~1, but it appeared a better fit can be made with the two-component depolarization function

\begin{multline}
A (t)= A_{tot}[f_{SC} \exp (-\frac{1}{2} \sigma_{SC}^2 t^2)\cos (2\pi\nu_{SC} t+\phi_{SC})
\\ + f_{N} \exp (-\frac{1}{2} \sigma_{N}^2 t^2)\cos (2\pi\nu_N t+\phi_N)] .
\end{multline}

\begin{figure}[ht]
\centering
\includegraphics[width=7cm]{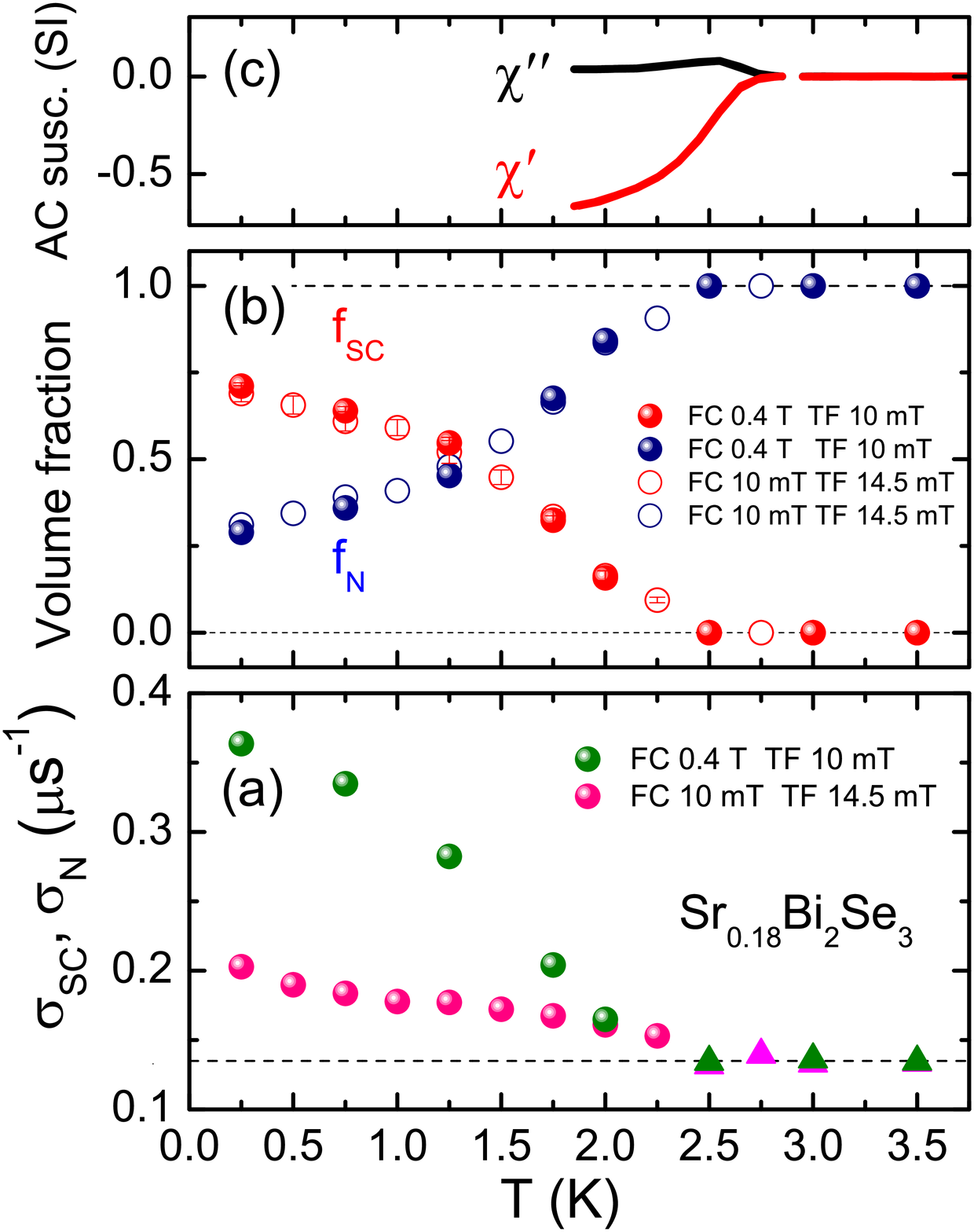}
\caption{Fit parameters of the two-component analysis (Eq.~4) of TF $\mu$SR spectra for Sr$_{0.18}$Bi$_{2}$Se$_{3}$. Disorder in the vortex lattice is induced by changing the field below $T_c$. Panel (a): $\sigma_{SC}(T)$ (round symbols) and $\sigma_{N}(T)$ (triangles). Green symbols: field-cooling in 0.4~T, spectra measured after reducing the field to TF = 10~mT ($B \parallel c$). Magenta symbols: cooling in 10~mT, spectra measured after sweeping the field first to zero and then up to TF= 14.5~mT ($B \parallel a$). Panel (b): Superconducting $f_{SC}$ and normal-state $f_N$ volume fraction for cooling in 0.4~T (closed symbols) and 10 mT (open symbols). Panel (c): AC susceptibility in S.I. units. The superconducting screening fraction is 0.7.}
\label{figure:2compfitpar}
\end{figure}

Here $f_{SC} = A_{SC}/A_{tot}$ and $f_{N} = A_{N}/A_{tot}$ are the volume fractions related to the superconducting and normal phases, respectively ($\nu_{SC}$, $\phi_{SC}$ and $\nu_{N}$, $\phi_{N}$ are the corresponding frequencies and phases). In the normal state $f_{SC} = 0$ and the relaxation rate $\sigma_N $ equals $0.134 \pm 0.002~\mu$s$^{-1}$. This value is close to the one reported in Fig.~\ref{figure:sigmaTF}. The result of the two-component fit of the spectrum at 0.25~K is shown in Fig.~\ref{figure:TFspectra_disorder}. Here the total asymmetry $A_{tot} = A_{SC}+A_{N}$ is fixed at the experimental value $A(0)=0.19$ (spin-rotated mode) and $\sigma_N$ is fixed at $0.134~\mu$s$^{-1}$. We obtain $f_{SC} = 0.71$, $f_N =0.29$ and $\sigma_{SC} = 0.36 \pm 0.02~\mu$s$^{-1}$. It shows that for this crystal the superconducting volume fraction amounts to 70~\%, in good agreement with ac-susceptibility measurements, see Fig.~\ref{figure:2compfitpar}(c). In Fig.~\ref{figure:2compfitpar} we show the temperature variation of the fit parameters $f_{SC}$, $f_N$ and $\sigma_{SC}$. The smooth variation of $f_{SC}$ to zero indicates $T_c =2.5$~K, which is slightly below the onset temperature for superconductivity $T_c =2.7$~K determined by ac-susceptibility on a piece of the same single-crystalline batch.

The values of $\sigma_{SC}$ in Fig.~\ref{figure:2compfitpar} indicate considerable disorder in the flux line lattice. In a second run we have field-cooled the sample in 10~mT to 0.25~K, next reduced the field to zero and subsequently increased it to 14.5~mT. TF field spectra ($B \parallel a$) taken after this field history showed $\sigma_{TF} = 0.20~\mu$s$^{-1}$ at 0.25~K, which indicates a much weaker degree of disorder in the vortex lattice. The temperature variation of the fit parameters $\sigma_{SC}$, $f_{SC}$ and $f_N$ obtained by using Eq.~4 for this second run are shown in Fig.~\ref{figure:2compfitpar}.

\section{DISCUSSION}

An important conclusion that can be drawn from the TF $\mu$SR spectra taken in the disordered vortex lattice case is that Sr$_{x}$Bi$_{2}$Se$_{3}$ for $x=0.18$ is a bulk superconductor. We remark that specific heat experiments around $T_c$, which provide a thermodynamic way to demonstrate bulk superconductivity, have not been reported in the literature so far. The superconducting volume fraction of 70~\% obtained by $\mu$SR nicely agrees with the superconducting screening fraction determined by ac-susceptibility measurements.

In the field-cooled case (ordered vortex lattice) we could not detect the damping of the $\mu^+$ precession signal due to superconductivity. This puts a lower bound on the penetration depth $\lambda$ of 2.3~$\mu$m. Within the London model $\lambda$ is related to the superfluid density $n_s$ via the relation
\begin{equation}
\lambda = (\frac{ m^*}{\mu _0 n_s e^2})^{1/2},
\end{equation}
where $m^*$ is the effective mass of the charge carriers, $\mu_0$ is the permeability of the vacuum and $e$ the elementary charge. Assuming $m^* = m_e$, $\lambda = 2.3~\mu$m translates to an extremely small value $n_s \sim 0.05 \times 10^{26}$~m$^{-3}$. This is difficult to reconcile with the carrier density $n = 1.2 \times 10^{26}$~m$^{-3}$ we measured by the Hall effect on a crystal from the same batch at 4.2~K. In the literature, however, significant lower values for $n$ have been reported: $0.27 \times 10^{26}$~m$^{-3}$ (Ref.~\onlinecite{Liu2015}) and $0.19 \times 10^{26}$~m$^{-3}$ (Ref.~\onlinecite{Shruti2015}), which results in $\lambda$-values of 1.0~$\mu$m and 1.2$~\mu$m, respectively, using Eq.~5. A possible solution is that $m^* > m_e$, but this is not in accordance with quantum oscillation studies. For low-carrier density samples of Bi$_{2}$Se$_{3}$ Shubnikov - de Haas data ($B \parallel c$) show $m^* =0.124 m_e$~\cite{Koehler1973b}. Doping may result in a slightly larger value of $m^*$. For instance for Cu-doped Bi$_{2}$Se$_{3}$ $m^* =$ 0.2-0.3$m_e$~\cite{Lahoud2013}. On the other hand, from specific heat experiments on Cu-doped Bi$_{2}$Se$_{3}$ a quasiparticle mass of $2.6m_e$ has been deduced~\cite{Kriener2011a}. Values for the effective mass of Sr$_{x}$Bi$_{2}$Se$_{3}$ have not been reported so far.

Very recently TF field muon spin rotation experiments on Cu$_{x}$Bi$_{2}$Se$_{3}$ crystals have been reported for a field of 10~mT applied along the $c$-axis~\cite{Krieger2017}. Interestingly, the authors do find a small increase of $\sigma_{TF}$ below $T_c$. In the normal state $\sigma_{TF} = 0.105 \pm 0.001 ~\mu$s$^{-1}$, a value comparable to the ones for the Sr doped case reported in Fig.~\ref{figure:sigmaTF}. In the superconducting phase a small but clear increase of $\sigma_{TF}$ is observed to a value of $0.113 \pm 0.001 ~\mu$s$^{-1}$ for $T \rightarrow 0$. By analyzing the data with help of Eq.~2 the authors calculate $\sigma_{SC}= 0.04~\mu$s$^{-1}$ and $\lambda = 1.6~\mu$m. We remark the total increase in $\sigma_{TF}$ below $T_c$ is only $0.008~\mu$s$^{-1}$, which is only slightly larger than the scatter in our values of $\sigma_{TF}$ (see Fig.~\ref{figure:sigmaTF}). The higher precision in these experiments is partly due to very long counting times resulting in better statistics. The $\mu$SR experiments on Cu and Sr doped Bi$_{2}$Se$_{3}$ agree in the sense that for both materials $\lambda$ is very large. Note that for Cu$_x$Bi$_{2}$Se$_{3}$ we calculate, with Eq.~5, using $\lambda = 1.6~\mu$m and assuming $m^* = m_e$ a superfluid density $n_s = 0.11 \times 10^{26}$~m$^{-3}$, which is also at variance with the measured carrier concentration~\cite{Kriener2011a,Lahoud2013} ($n_s$ is a factor 10 smaller). The recurring result that $n_s \ll n$ seems to indicate that only part of the conduction electrons participate in the superconducting condensate. A possible explanation is substantial electronic phase inhomogeneities, where the superconducting phase (volume fraction 70 \% for Sr$_{0.18}$Bi$_{2}$Se$_{3}$ and 40-60~\% for Cu$_x$Bi$_{2}$Se$_{3}$~\cite{Kriener2011a,Krieger2017}) has effectively a lower carrier concentration than the normal phase. On the other hand, a similar mismatch between $n_s$ and $n$ has recently been reported for the Nb doped low-carrier density superconductor SrTiO$_3$ notably in the over-doped, dirty regime, which is relevant in the context of high-T$_c$ cuprates as well~\cite{Collignon2017}. We remark that the discrepancy between $n_s$ and $n$ does not show up in the standard analysis of the Ginzburg-Landau parameter $\kappa = \lambda /\xi$, where $\xi$ is the superconducting coherence length. The large value of $\kappa \sim 100$ and the small coherence length $\xi \sim 15$~nm extracted from transport and magnetic measurements~\cite{Shruti2015,Kriener2011a} result in a substantial value $\lambda \sim 1.5$~$\mu$m. Here we have neglected for the purpose of simplicity the crystalline anisotropy of about a factor 1.5 in these parameters.

The $\mu$SR spectra for the $x=0.18$ crystal, taken after cooling in 0.4~T and subsequently reducing the field to 10~mT, show a sizeable depolarization due to disorder in the vortex lattice. If we assume a random distribution of flux lines, $\lambda$ can be calculated using the expression $\langle
(\Delta B)^2 \rangle = \Phi _0 B/ 4 \pi \lambda ^2$ (see Refs~\onlinecite{Belousov1990,Aegerter&Lee1997}). With $\sigma_{SC} = 0.36~\mu$s$^{-1}$ (see Fig.~\ref{figure:2compfitpar}(a)) we calculate $\lambda = 3.0~\mu$m, a value in line with the lower bound 2.3~$\mu$m estimated from the field-cooled experiments. It is not surprising substantial disorder in the vortex lattice can be created. In the Cu, Sr and Nb case experimental evidence has been presented that the dopant atoms are intercalated in the Van der Waals gaps between the quintuple layers of the Bi$_{2}$Se$_{3}$ structure~\cite{Hor2010,Liu2015,Qiu2015}. However, partial substitution on the Bi lattice cannot be ruled out. A detailed refinement of the crystal structure after intercalation has not been reported for these compounds so far. For Cu$_x$Bi$_{2}$Se$_{3}$ it has been inferred by analogy to related selenides that the intercalant atoms reside in the $3b$ site (Wyckoff notation)~\cite{Hor2010}. Moreover, structural investigations report considerable disorder on various length scales~\cite{Kriener2011b,Mann2014}. Thus the Bi-based superconductors are prone to various types of structural disorder, which in turn may provide different sources of flux pinning.

\section{Summary and conclusions}

We have performed transverse field muon spin rotation experiments on single-crystalline samples of Sr$_{x}$Bi$_{2}$Se$_{3}$ with the aim to investigate the crystalline anisotropy in the London penetration depth, $\lambda$. Field-cooled $\mu$SR spectra measured for the ordered flux line lattice reveal however no additional damping of the $\mu ^+$ precession signal in the superconducting phase. From the data we infer a lower bound for $\lambda$ of $2.3~\mu$m. By changing the applied magnetic field in the superconducting phase we are able to induce disorder in the vortex lattice. This results in a sizeable value $\sigma _{SC} = 0.36 ~\mu$s$^{-1}$ for $T \rightarrow 0$. By analyzing the $\mu$SR time spectra with a two component function we obtain a superconducting volume fraction of 70~\%. This provides solid evidence for bulk superconductivity in Sr$_{x}$Bi$_{2}$Se$_{3}$. We signal a discrepancy between the superfluid density, $n_s$, calculated from $\lambda$ within the London model, and the measured carrier concentration. An important question that has not been answered yet is the two-fold basal-plane anisotropy of $\lambda$. This asks for $\mu$SR experiments with improved statistics or another technique to probe $\lambda$, for instance, a tunnel diode oscillator circuit. Finally, we recall that the reported~\cite{Pan2016,Matano2016,Yonezawa2017,Asaba2017} breaking of rotational symmetry in the small family of Bi$_2$Se$_3$-based superconductors deserves a close examination, notably because it offers an excellent opportunity to study unconventional superconductivity with a two-component order parameter.

\
\textbf{Acknowledgements}
\
This work was part of the research program on Topological Insulators funded by FOM (Dutch Foundation for Fundamental Research of Matter). The research of H. Leng was made possible by grant 201604910855 of the Chinese Scholarship Council. The authors are grateful to M.J. Graf for help with the experiments and fruitful discussions. Part of this work was performed at the Swiss Muon Source (S$\mu$S) of the Paul Scherrer Institute, Villigen, Switzerland.

\bibliography{Refs_SrBi2Se3}

\bibliographystyle{apsrev4-1}

\end{document}